\documentstyle[aps,preprint,epsf]{revtex}

\begin{document}
\tightenlines
\draft

\title{The $\nu \nu \gamma$ Amplitude in an External Homogeneous 
Electromagnetic Field}

\author{R.~ SHAISULTANOV\thanks{
Email:shaisul@pion14.tphys.physik.uni-tuebingen.de
}}

\address{Institut f\"{u}r Theoretische Physik, 
Universit\"{a}t T\"{u}bingen,\\
 Auf der Morgenstelle 14, 72076 T\"{u}bingen, Germany \\
and \\
Budker Institute of Nuclear Physics \\
630090, Novosibirsk 90, Russia}

\maketitle
\begin{abstract}
Neutrino-photon interactions in the presence of an external 
homogeneous constant electromagnetic field are studied. 
The $\nu \nu \gamma $ amplitude is calculated in an electromagnetic 
field of the general type, when the two field invariants are nonzero.
\end{abstract}
 \newpage
\narrowtext
\section{Introduction}
Neutrino-photon interactions are of interest in astrophysics
and cosmology. It is well known that in the standard model, 
neutrino-photon interactions appear at the one-loop level. 
It is also well known that in a vacuum the reaction 
$\nu \rightarrow \nu \gamma$ is forbidden. The processes with 
two photons, for example, neutrino-photon scattering 
\( \gamma \, \, \nu \rightarrow \gamma \, \, \nu \, \,  \), 
turn out to be highly suppressed. In \cite{a}
Gell-Mann showed that in the four-Fermi limit of the standard 
model the amplitude
is exactly zero to order \( G_{F} \) because, according to 
Yang's theorem \cite{yang}, two
photons cannot couple to the \( J=1 \) state. Therefore the 
amplitude is suppressed
by the additional factors of \( \omega /m_{W} \), where \( \omega  \) 
is the
photon energy and \( m_{W} \) is the \( W \) mass \cite{a,a2,b,c}. 
For example,
in the case of massless neutrinos, the amplitude for 
\( \gamma \nu \rightarrow \gamma \nu  \)
in the standard model is suppressed by the factor 
\( 1/m_{W}^{2} \) \ \cite{f}
. As a result, the cross section is exceedingly small. 
So we see that neutrino-photon processes with one or two 
photons do not occur or are suppressed in the vacuum.

The presence of a medium or external electromagnetic field 
drastically changes the situation. It induces an effective 
coupling between photons and neutrinos,
which contributes to the  $\nu \rightarrow \nu \gamma$ process 
and cross-related reactions. Furthermore, it was shown in 
\cite{rashid} that in the presence of an 
external magnetic field, cross sections for neutrino-photon processes 
such as $\gamma\gamma\to \nu\bar{\nu}$ and
       $\nu\gamma\to \nu\gamma$ are enhanced by the factor
\( \sim \left( m_{W}\, \, /m_{e}\right) ^{4}\left( B/B_{c}\right) ^{2} \) 
for $  \omega \ll m_{e}$ and $B \ll B_{c}$.
Later this result was extended to very strong
magnetic fields \cite{vas} and arbitrary $\omega$ \cite{kit}. 

In this paper we will deal with $\nu \rightarrow \nu \gamma$ and 
$ \gamma \rightarrow \nu \, \, \overline{\nu } $  reactions in 
the presence of
an external electromagnetic field. The photon decay process 
$\gamma\rightarrow \nu \, \, \overline{\nu }$ in presence of a 
magnetic field was studied by several authors \cite{gal,milton}. 
The Cherenkov process $\nu \rightarrow \nu \gamma $ is the cross 
process to photon decay and was studied in a crossed field \cite{gal} 
and in a magnetic field in \cite{skobel,raf}.
The aim of this paper is to consider the case of an arbitrary
homogeneous electromagnetic field. This case was also considered
recently in \cite{sy} but, from our point of view, the expression
obtained is unfinished insofar as the very short tensorial structure
of the V-A two-point amplitude, which we derive in (\ref{ha9}), is
nowhere visible in \cite{sy}. Therefore it was not possible to compare
our expressions to Schubert's formula (5.15). Since this topic abounds 
in errors and controversies ( see \cite{raf} for a thorough discussion 
of the literature and its critique), we will devote this article mainly 
to a careful analysis of this expression and will postpone physical 
applications to another forthcoming paper.

\section{ The $\nu \nu \gamma$ Process in Presence of External Fields}

Let us begin with the effective Lagrangian. At energies very much smaller
 than the W- and Z-boson masses, both processes are described by an 
 effective four-fermion interaction,
\begin{equation}
\label{ha1}
{ \cal L } _{eff}=\frac{G_{F}}{\sqrt{2}}\overline{\nu }
\gamma ^{\mu }\left( 1+\gamma _{5}\right) \nu \overline{E}
\gamma _{\mu }\left( g_{V}+g_{A}\gamma _{5}\right) E.
\end{equation}

Here, $E$ stands for the electron field,
$\gamma_5=-i\gamma^0\gamma^1\gamma^2\gamma^3$,  
$g_V=1-\frac{1}{2}\left(1-4\sin^2\theta_W\right)$ and 
$g_A=1-\frac{1}{2}$ for $\nu_{e}$ , where the first terms 
in $g_V$ and $g_A$ are the contributions from
the W exchange diagram and the second one from the Z 
exchange diagram. Also,
$g_V=2\sin^2\theta_W-\frac{1}{2}$ and
$g_A=-\frac{1}{2}$ for $\nu_{\mu,\tau}$. Then the amplitude 
for the diagram in Fig.1 is given by
\begin{equation}
\label{m}
{\cal M}=\frac{G_F}{\sqrt{2}\,e}\varepsilon_{\mu}
\bar{\nu}\gamma_{\nu}(1+\gamma_5)\nu\,
(g_V\Pi^{\mu \nu}+g_A\Pi_5^{\mu \nu})\, ,
\end{equation} 
where $ \Pi^{\mu \nu}$ is the well-known polarisation operator of QED. 
It was calculated earlier in \cite{shabad,tsai0,ditt,bks0,urr}. 
Our aim is to calculate $\Pi_5^{\mu \nu}$. For this purpose we will 
use the technique developed in \cite{sch,tsai,tsai1,bks,bks0}. 
In this approach we begin with
\begin{equation}
\label{ha2}
 \Pi_5^{\mu \nu} =-i e^2 M_5^{\mu \nu},
\end{equation}
where 
\begin{equation}
\label{ha3}
M_5^{\mu \nu}\equiv Sp\left\langle 0\right| \gamma ^{\mu }\frac{1}
{{\cal P}\!\!\!\!/-m+i\varepsilon }\gamma ^{\nu }\gamma ^{5}
\frac{1}{{\cal P}\!\!\!\!/-k\!\!\!/-m+i\varepsilon }\left| 0\right\rangle ,
\end{equation}
with ${\cal P}_{\mu}=i {\partial}_{\mu}-e A_{\mu}$, and we have to 
calculate the mean value over the states x=0 : 
$\langle 0\left|\ldots\right| 0\rangle $. For the analysis it is 
convenient to use a special representation for the field tensor:
\begin{equation}
\label{ha6}
F_{\mu \nu }=aC_{\mu \nu }+bB_{\mu \nu }\, ,
\end{equation}
with the tensors $C_{\mu \nu }$ and $B_{\mu \nu }$ defined by 
\begin{equation}
\label{ha65}
C_{\mu \nu }=\frac{\displaystyle 1}{\displaystyle a^{2}+b^{2}}
\left( aF_{\mu \nu }+bF^{*}_{\mu \nu }\right)\, ;   
B_{\mu \nu }=\frac{\displaystyle 1}{\displaystyle a^{2}+b^{2}}
\left( bF_{\mu \nu }-aF^{*}_{\mu \nu }\right)\, , 
\end{equation}
where 
\begin{equation}
\label{ha4}
a,b=\sqrt{({\cal F}^{2}+{\cal G}^{2})^{\frac{1}{2}}
\pm {\cal F}}\,\,;\,\,\, {\cal F}=-\frac{1}{4}F_{\mu \nu }
F^{\mu \nu }\, ,\,\,{\cal G}=-\frac{1}{4}F^{*}_{\mu \nu }F^{\mu \nu } .
\end{equation}
Then\footnote{We always use
the metric $g={\rm diag}({+}{-}{-}{-})$.}
\begin{equation}
\label{ha655}
\left( C^{2}\right) _{\mu \nu }=\frac{\displaystyle 1}
{\displaystyle a^{2}+b^{2}}\left( F^{2}_{\mu \nu }+b^{2}
g_{\mu \nu }\right)\, ;\;\;  \left( B^{2}\right) _{\mu \nu }=
\frac{\displaystyle 1}{\displaystyle a^{2}+b^{2}}
\left( F^{2}_{\mu \nu }-a^{2}g_{\mu \nu }\right)\, .
\end{equation}
These tensors satisfy very useful identities:
\begin{equation}
\label{ha657}
\left( C B \right)_{\mu \nu}=0 \, ;\;\;\;\;\left( C^{3} \right)_{\mu \nu}=
C_{\mu \nu} \;;\;\;  \left( B^{3} \right)_{\mu \nu}=-B_{\mu \nu} \,. 
\end{equation}

Now the result of our calculation is given by
\begin{eqnarray}
\label{ha7}
M^{\mu \nu }_{5} & =2i\frac{\displaystyle \pi ^{2}}
{\displaystyle (2\pi )^{4}}\{2ebC^{\mu \nu }-2eaB^{\mu \nu }+
\displaystyle \int ^{1}_{-1}dv\int ^{\infty }_{0}sds
\frac{\displaystyle ie^{2}ab}{\displaystyle \sin (ebs)\sinh (eas)}
\exp [i\Psi ]e^{-ism^{2}} & \nonumber \\
 & [(Ck)^{\mu }(C^{2}k)^{\nu } M_{1}+(Bk)^{\mu }(B^{2}k)^{\nu } M_{2}+ & \\
 & [(C^{2}k)^{\mu }(Bk)^{\nu }+(Bk)^{\mu }(C^{2}k)^{\nu }+
 (kC^{2}k)B^{\mu \nu }] M_{3} & \nonumber \\
 & [(B^{2}k)^{\mu }(Ck)^{\nu }+(Ck)^{\mu }(B^{2}k)^{\nu }+
 (kB^{2}k)C^{\mu \nu }] M_{4}]\}, & \nonumber \\\nonumber
\end{eqnarray}
where
\begin{equation}
\label{ha8}
\Psi =\frac{(kC^{2}k)}{2ea}\frac{\cosh (eas)-\cosh (easv)}
{\sinh (eas)}+\frac{(kB^{2}k)}{2eb}\frac{\cos (ebs)-\cos (ebsv)}{\sin (ebs)},
\end{equation}
with
\begin{eqnarray}
\label{m1}
M_{1} & =&  \sin (ebs)(\cosh (eas)-\cosh (easv))\frac{\displaystyle 1}
{\displaystyle \sinh ^{2}(eas)}  \\
M_{2} & = & \sinh (eas)(-\cos (ebs)+\cos (ebsv))\frac{\displaystyle 1}
{\displaystyle \sin ^{2}(ebs)} \nonumber \\
M_{3} & = & \frac{\displaystyle 1}{\displaystyle 2}(-\cos (ebsv)
\cosh (easv)\coth (eas)+\cos (ebsv)\frac{\displaystyle 1}
{\displaystyle \sinh (eas)}+\nonumber \\
&&\cot (ebs)\sin (ebsv)\sinh (easv)) \nonumber \\
M_{4} & = & \frac{\displaystyle 1}{\displaystyle 2}
(\cos (ebsv)\cosh (easv)\cot (ebs)-\cosh (easv)
\frac{\displaystyle 1}{\displaystyle \sin (ebs)}+\nonumber \\
&&\coth (eas)\sin (ebsv)\sinh (easv))\, . \nonumber
\end{eqnarray}

%\begin{equation}
%Q^{\mu \nu }=2bC^{\mu \nu }-2aB^{\mu \nu }\end{equation}

As in the previous calculations in presence of a magnetic field 
\cite{milton,raf}, this result is not gauge invariant. We also see 
that our amplitude is ultraviolet convergent. Nevertheless, in
order to preserve gauge invariance, we must regularize it using a 
gauge invariant regularization. This phenomenon should be familiar 
from quantum electrodynamics. For example, the photon-photon scattering 
amplitude, though formally converging, must be regularized to get a 
gauge invariant result \cite{lif,jackiw}. A convenient way of 
restoring gauge invariance in diagrams with pseudovector coupling 
is to use the Pauli-Villars regularization. Then it is easy to show 
that to obtain the final gauge invariant expression for the amplitude 
we must just omit the first two terms in (\ref{ha7}), because they are 
cancelled by regulator terms. Finally, we obtain
\begin{eqnarray}
\label{ha9}
M_{5}^{\mu \nu } & = & -\frac{\displaystyle 1}{\displaystyle 8\pi ^{2}}
\displaystyle \int ^{1}_{-1}dv\int ^{\infty }_{0}sds\frac{\displaystyle 
e^{2}ab}{\displaystyle \sin (ebs)\sinh (eas)}\exp [i\Psi ]e^{-ism^{2}}
\left\{ \right. (Ck)^{\mu }(C^{2}k)^{\nu }M_{1}+ \nonumber \\
 &  & (Bk)^{\mu }(B^{2}k)^{\nu }M_{2}+[(C^{2}k)^{\mu }(Bk)^{\nu }+
 (Bk)^{\mu }(C^{2}k)^{\nu }+(kC^{2}k)B^{\mu \nu }]M_{3}+\\
 &  & [(B^{2}k)^{\mu }(Ck)^{\nu }+(Ck)^{\mu }(B^{2}k)^{\nu }+
 (kB^{2}k)C^{\mu \nu }]M_{4}\left. \right\} .\nonumber 
\end{eqnarray}

 This method of getting a gauge invariant expression allows us, 
unlike the authors of \cite{raf}, to do without an anomaly 
cancellation mechanism at this stage of calculation. Let us 
note that the whole contribution of the triangle diagram is 
still present in (\ref{ha9}). This fact can be easily confirmed 
by an independent calculation of the triangle graph contribution. 
To deal with unwanted anomaly terms we must, as usual, take into 
account all fermions in a generation. In the following this  will 
always be assumed.

Equation (\ref{ha9}) is the main result of this article. 
Let us now consider
various checks  of this expression. First, in the limit of zero 
electric field it reproduces the results of \cite{milton,raf}. 
A second nontrivial check of formula  (\ref{ha9}) is to consider 
the following transformation: 
\begin{equation}
\label{m2}
a \rightarrow i b ; b \rightarrow -i a ; C_{\mu \nu} 
\rightarrow  -i B_{\mu \nu}; B_{\mu \nu} 
\rightarrow  i C_{\mu \nu}; F_{\mu \nu} \rightarrow F_{\mu \nu}\, . 
\end{equation}
Because of  $ F_{\mu \nu} \rightarrow F_{\mu \nu}$, $ F^{*}_{\mu \nu} 
\rightarrow F^{*}_{\mu \nu}$, the tensor $M_{5}^{\mu \nu }$ must be 
invariant under this transformation. From (\ref{ha9}) we see that it 
is indeed invariant and that under this transformation, the first 
term is interchanged with the second, and the third term is 
interchanged with the fourth, thus demonstrating the nontriviality 
of this check.

The next interesting limit is the case of a crossed field, where 
$\vec{E}$ and $\vec{B}$ are not only orthogonal but are also of 
equal modulus. This causes the two invariants $\cal{F}$ and $\cal{G}$ 
to vanish. The field dependence is therefore completely described 
by the invariant $( F^{\mu \nu} k_{\nu})^{2}$ . The correct procedure 
of finding the amplitude is to set $a=b$ first, and then take the 
limit $a=b \rightarrow 0$. Using this procedure we find following 
expression:
\begin{eqnarray}
\label{m3}
M^{\mu \nu }_{5} & = &-  \frac{e}{8\pi ^{2}}\left\{ \frac{1}{2m^{2}}
\left[ k^{2}F^{*\mu \nu }+(F^{*}k)^{\mu }k^{\nu }+(F^{*}k)^{\nu }k^{\mu }
\right] \right. \nonumber \\
 &  & \int _{0}^{1}dv\left( 1-v^{2}\right) u\int _{0}^{\infty }dz
 \exp (-i(zu+\frac{z^{3}}{3}))+\\
 &  & \left. \frac{4}{(kF^{2}k)}(F^{*}k)^{\mu }(F^{2}k)^{\nu }
 \left\{ \int _{0}^{1}dvu\int _{0}^{\infty }dz
 \exp (-i(zu+\frac{z^{3}}{3}))+i\right\} \right\}\, , \nonumber
\end{eqnarray}
where
\begin{equation}
\label{m4}
u=\left( \frac{e^{2} (kF^{2}k)}{16m^{6}}\right) ^{-\frac{1}{3}}
(1-v^{2})^{-\frac{2}{3}} \, .
\end{equation} 
Earlier the amplitudes in crossed fields were considered in 
\cite{gal,mik}.
We can compare only parts of (\ref{m3}) with those existing in 
the literature, because 
all previous results were presented in contracted form and with 
photons on the mass shell. Taking this into account we can say 
that our result contradicts that of \cite{gal} and confirms 
that of \cite{mik}.  
\section{Conclusions}

We have presented the $\nu \nu \gamma$ amplitude in the presence of an
 external homogeneous constant electromagnetic field. The result is
 gauge invariant and reproduces the known results for an external
 magnetic field. We further identified the crossed-field limit. Let us
 also emphasize that the general case of electric and magnetic fields
 is in need when considering ( low-frequency ) multiphoton one-loop
 processes with and without external fields. The situation is similar
 to processes where the Heisenberg-Euler lagrangian turns out to be
 useful as e.g., in the study of photon-splitting \cite{al} or in
 photon-neutrino processes as discussed in the recent article by Dicus
 and Repko \cite{dr}. Hence we conclude that
 our results allow for a variety of applications that will also be 
 discussed in a future publication.

\section*{Acknowledgments}

I would like to thank Professor W. Dittrich for helpful
discussions and for carefully reading the manuscript. I am also
grateful to Dr H. Gies for valuable comments and suggestions.
This work was supported by Deutsche Forschungsgemeinschaft
under DFG Di 200/5-1.

\begin{figure}
\centerline{\epsfysize=0.9\textheight \epsffile{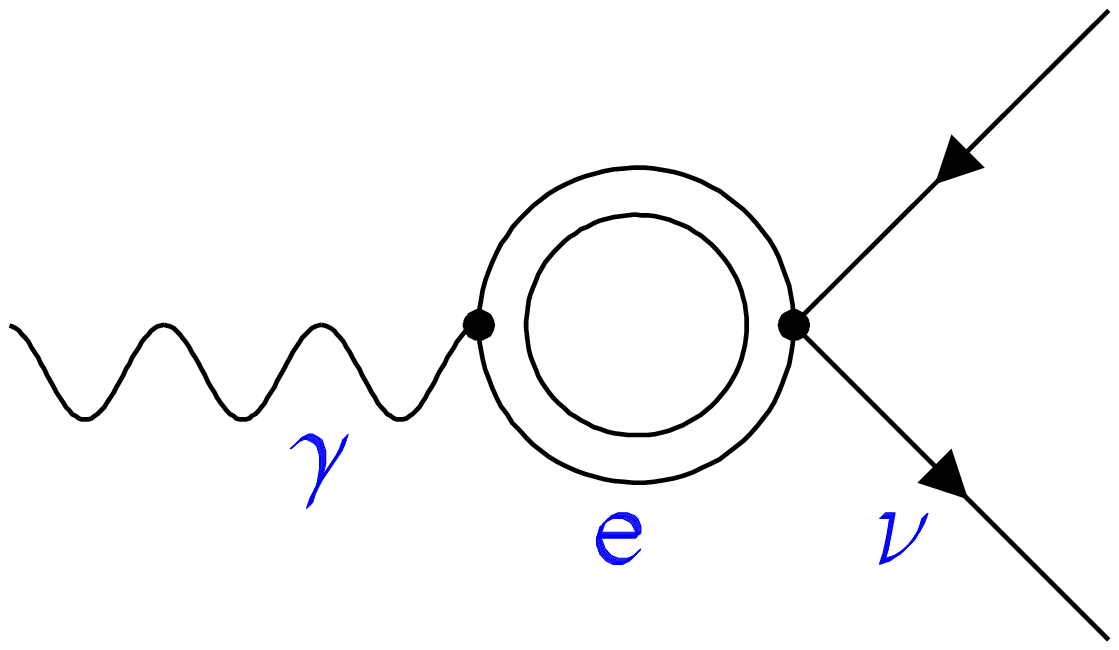}}
\caption{Neutrino-photon process in an external electromagnetic field.
The double line represents the electron propagator in the presence of
an external field.}
\end{figure}

\end{document}